\documentclass[aps,prb,twocolumn,showpacs,preprintnumbers]{revtex4}
\usepackage{graphicx} 
\usepackage{amsmath}
\usepackage{graphicx,epstopdf}
\usepackage{graphicx}
\usepackage{epstopdf}
\usepackage{mathtools}
\usepackage{tikz}
\usepackage{color}
\usetikzlibrary{shapes.misc,shadows}

\newcommand{\be}{\begin{equation}}
\newcommand{\ee}{\end{equation}}
\newcommand{\bea}{\begin{eqnarray}}
\newcommand{\eea}{\end{eqnarray}}
\newcommand{\bse}{\begin{subequations}}
\newcommand{\ese}{\end{subequations}}

\begin{document}

\title {Velocity auto correlation function of a confined Brownian particle}

\author{Arsha N}
\author{Shabina S}
\author{Mamata Sahoo}
\email{jolly.iopb@gmail.com}
\affiliation{Department of Physics, University of Kerala, Kariavattom, Thiruvananthapuram-$695581$, India}
\date{\today}

\begin{abstract}
Motivated by the simple models of molecular motor obeying a linear force-velocity relation, we have studied the stochastic dynamics of a Brownian particle in the presence of a linear velocity dependent force, $f_s(1-\frac{v}{v_0})$ where $f_{s}$ is a constant. The position and velocity auto correlation functions in different situations of the dynamics are calculated exactly. We observed that the velocity auto correlation function shows an exponentially decaying behaviour with time and saturates to a constant value in the time asymptotic limit, for a fixed $f_s$. It attains saturation faster with increase in the $f_{s}$ value. When the particle is confined in a harmonic well, the spectral density exhibits a symmetric behaviour and the corresponding velocity auto correlation function shows a damped oscillatory behaviour before decaying to zero in the long time limit. With viscous coefficient, a non-systematic variation of the velocity auto correlation function is observed. Further, in the presence of a sinusoidal driving force, the correlation in velocities increases with increase in the amplitude of driving in the transient regime. For the particle confined in a harmonic well, the correlation corresponding to the shift relative to the average position is basically the thermal contribution to the total position correlation. Moreover, in the athermal regime, the total correlation is entirely due to the velocity dependent force.
\end{abstract}
\pacs{75.50.Ee, 71.20.Ps, 75.10.Pq, 75.30.Kz, 75.30.Et, 75.10.Jm}
\maketitle

\section {\textbf{Introduction}}
Stochasticity is present in almost all the processes occurring in nature encompassing biological processes, processes occurring at nano scale, financial marketing, and human organizations\cite{strogatz1994nonlinear}. Generally, stochastic processes are very complex in nature because of the presence of multitude interactions. Brownian motion is one of the simplest example of a stochastic system\cite{einstein2005theorie}. It is basically the random movement of a particle placed in a colloidal solution, where the size of the particle is comparatively larger than the size of the colloidal suspension particles\cite{nelson1967dynamical,mazo2002brownian,babivc2005colloids,uhlenbeck1930theory,CHOW1973156,chandrasekhar1949brownian,
parisi2005brownian}. Over the years, the study of Brownian dynamics of a particle in the microscopic environment has become an active area of research in both Physics and Biology\cite{frey2005brownian,Astumian917,squires2010fluid, wirtz2009particle,hanggi2009artificial}.

Langevin's theory is typically used in modelling the stochastic processes in a Brownian environment \cite{langevincomptes,tothova2011langevin}. For instance, when the motion of a microscopic particle is in contact with a heat bath or Brownian environment, the Langevin's model mainly helps in describing the interaction of the particle with the Brownian environment. Due to this interaction, the environment exerts two different kinds of forces on the particle, i.e., the viscous force and the random force or noise. The viscous force always try to slow the motion of the particle and acts in a direction opposite to it's motion. On the other hand, the random force changes it's direction and magnitude frequently compared to any other time scales involved in the process and becomes zero when averaged over time\cite{kubo2012statistical}.
In contrast to the dynamics of a conventional Brownian particle, where the forces are due to the external sources, the active Brownian particles can generate the forces by their own, known as the self propulsion forces which can be velocity dependent\cite{howse2007self,reimann2002brownian,paxton2004catalytic,hagen2011brownian,schweitzer2003brownian,schweitzer1998complex,
zhang2008active}. 
 
Generally, molecular motors such as kinesins move on the molecular track in a highly stochastic but directed manner, making use of the chemical energy from ATP hydrolysis. If a load force is applied in a direction opposite to their motion, the motor gets  slowed down and eventually stop moving. One can model this behaviour of the molecular motor using a linear force-velocity relation \cite{howard2001mechanics, savoda1994cell,ganguly2013}. For instance, if we assume $f_{a}$ as the autonomous force generated by the molecular motor, then in the presence of an external load force, say $\lambda$, the dynamics can be described by Langevin's equation of motion $\dot{v}=-\gamma v+f_{a}-\lambda+\xi(t)$. This leads to a linear force velocity relation, $<v>=v_{0}(1-\lambda/f_a)$ in the over damped limit, with $v_{0}=\frac{f_{a}}{\gamma}$ as the autonomous velocity of the free motor, $\xi(t)$ as the random force and $f_{a}$ as the stall force. If we replace $f_{a}$ by a linear velocity dependent force $f_{a}(1-\frac{v}{v_{0}})$, it will modulate the effective viscous  coefficient by a constant additive amount. This change of viscous drag can be thought of as the mechanochemical processes leading to self propulsion in case of molecular motors. 

Motivated by this behaviour of molecular motor, in this work, we have studied the dynamics of a Brownian particle in the presence of a linear velocity dependent force by analyzing the position and velocity auto correlation functions. As the position and velocity autocorrelation functions are related to the mean square displacement or diffusion and mean kinetic energy of the particle, respectively, the dynamical behaviour as well as the complex interactions of a model system can be well understood by analyzing these two physical quantities. Using molecular dynamic simulation, the velocity autocorrelation function has been studied earlier in various contexts \cite{volpe2013simulation,Bellissima,chakraborty2011velocity,Williams,do2012force,wright2003forward,cichocki,ATZBERGER2006225,
hanna1981velocity}. 
In all these reports, the studies are mostly focused on the scaling behaviour of the velocity autocorrelation function in different time regimes. For instance, though, an exponential decay of the velocity autocorrelation function is predicted for a simple Langevin model, it is found that the decay exhibits three distinct time regimes representing three different states of the solvent or medium\cite{chakraborty2011velocity}. In another report, the time dependent velocity autocorrelation is investigated with varying density. The presence of distinct fast and slow decay channels for the velocity autocorrelation on the time scale set by the collision rate at various densities are identified\cite{Bellissima}. Herein, we have considered a simple Langevin model of the Brownian dynamics and exactly calculated the position and velocity autocorrelation functions in different parameter regimes of the model. We have discussed the influence of these model parameters on the time evolution of the dynamics under various circumstances.

\section{\textbf{Model}}
We have considered the dynamics of a Brownian particle in an effective potential $U(x,t)$ and in the presence of a linear velocity dependent force, $f_s(1-\frac{v}{v_0})$, with $f_{s}$ as constant and $v_{0}$ as the autonomous velocity of the particle, respectively \cite{ganguly2013}. The motion of the particle is governed by the Langevin equation of motion\cite{chandrasekhar1943stochastic,jayannavar2007charged,sahoo2019transport} 
\begin{equation}
m\frac{dv}{dt}=-\gamma v-\frac{\partial U\left( x,t\right) }{\partial x}+f_s\left( 1-\frac{v}{v_0}\right)+\xi\left(t\right).
\end{equation}
Here, $m$ and $v(t)$ represent the mass and instantaneous velocity of the Brownian particle, respectively and $\gamma$ is the viscous coefficient of the medium. $\xi(t)$ is the randomly fluctuating thermal noise which satisfies the properties, $<\xi(t)>=0$ and $<\xi(t_1)\xi(t_2)>=2\gamma k_BT\delta(t_{1}-t_{2})$. The angular bracket $\left\langle ...\right\rangle $ represents the ensemble average over the noise, $k_B$ is the Boltzmann constant, and $T$ is the absolute temperature of the medium or environment. The potential $U(x,t)$ depends on both position($x$) and time($t$).

In this work, we have studied the dynamics for three different cases. In all these studies, we are mainly focused on the nature of the position and velocity auto correlation functions. First, we have simply considered the inertial dynamics of a Brownian particle in the presence of a linear velocity dependent force, $f_s(1-\frac{v}{v_0})$. In this case, since the particle is not confined in a potential,
$U(x,t)=0$, and hence Eq.~(1) reduces to
\begin{equation}
m\frac{dv}{dt}=-\gamma v+f_s\left( 1-\frac{v}{v_0}\right)+\xi\left( t\right).
\end{equation}
By solving Eq.~(2), the velocity, $v(t)$ at any instant of time is found to be
\begin{equation}
\begin{split}
v\left( t\right) =v_0 e^{-\frac{\beta}{m} t}+\left( \frac{f_s}{m}\right) \int_{0}^{t}e^{-\frac{\beta}{m}\left( t-t^{'}\right)} dt^{'}+\\\frac{1}{m}\int_{0}^{t} e^ {-\frac{\beta}{m}\left( t-t^{'}\right)} \xi\left( t^{'}\right) dt^{'},
\end{split}
\end{equation}
where $\beta=(\gamma+\frac{f_s}{v_0})$ and $g=2\gamma k_{B}T$. Thus, the velocity auto-correlation function, $C_v(t_{1},t_{2})=\left[\left\langle v(t_{1})v(t_{2})\right\rangle \right]$ is calculated as
\begin{equation}
\begin{split}
 C_v(t_{1},t_{2})=v_{0}^{2}e^{\dfrac{-\beta (t_{1}+t_{2})}{m}}\\+\dfrac{v_{0}f_{s}}{\beta}\left(  e^{\dfrac{-\beta t_{1}}{m}}+e^{\dfrac{-\beta t_{2}}{m}}-2e^{\dfrac{-\beta (t_{1}+t_{2})}{m}}\right)\\+\left( \dfrac{f_{s}}{\beta} \right) ^{2}\left(1- e^{\dfrac{-\beta t_{2}}{m}}-e^{\dfrac{-\beta t_{1}}{m}}+e^{\dfrac{-\beta (t_{1}+t_{2})}{m}}\right)\\+\dfrac{g}{2m\beta}\left( e^{\dfrac{-\beta \vert(t_{1}-t_{2})\vert}{m}}-e^{\dfrac{-\beta (t_{1}+t_{2})}{m}}\right).
 \end{split}
\end{equation}
From this expression, the stationary state correlation function $C_{v}(t)(=<v(0)v(t)>)$ can be calculated as
\begin{equation}
C_{v}(t)=\left( \dfrac{f_{s}}{\beta} \right) ^{2}+\dfrac{g}{2m\beta}\left( e^{\dfrac{-\beta t}{m}}\right)
\end{equation}
Next, we have attempted to study the position and velocity autocorrelation functions when the particle is confined in a harmonic well $[U(x,t)=\frac{1}{2}kx^{2}]$. The dynamics of the particle is given by  
\begin{equation}
m\frac{dv}{dt}=-kx-\gamma v+f_s\left( 1-\frac{v}{v_0}\right)+\xi\left( t\right),
\end{equation}
where $k$ is the harmonic constant. From the solution of Eq.~(5), we get the position [$x(\omega)$] and velocity [$v(\omega)$] of the particle in the Fourier space as 
\begin{eqnarray}\nonumber
x\left( \omega\right) =\left(\frac{f_s\delta\left( \omega\right)+\xi \left( \omega\right) }{-m\omega^{2}-i \omega \beta +k}\right)
\end{eqnarray}
and
\begin{eqnarray}\nonumber
v\left( \omega\right) =\omega\left( \frac{f_s\delta\left( \omega\right)+\xi \left( \omega\right) }{-im\omega^{2}+\omega  \beta +ik}\right).
\end{eqnarray}
The spectral densities corresponding to the position $S_x\left( \omega\right)$ [=$\mid x\left( \omega \right) \vert\ ^{2} $] and velocity $S_v\left( \omega\right)$ [= $\mid v\left( \omega \right) \vert\ ^{2} $] are 
\begin{equation}
S_{x}\left( \omega\right) =\left(\frac{f_{s}^{2}+2 \gamma k_{B}T}{\left( m \omega^{2}-k\right) ^{2}+\left( \omega \beta\right) ^{2}}\right)
\end{equation}
and
\begin{equation}
 S_{v}\left( \omega\right) =\omega^{2}\left( \frac{ f_{s}^{2}+2 \gamma k_{B}T}{\left( m \omega^{2}-k\right) ^{2}+\left( \omega \beta\right) ^{2}}\right).
\end{equation}
By doing the inverse Fourier transform of Eqs.~(7) and (8), the position autocorrelation function, $C_x(t)$ and the velocity autocorrelation function $C_v(t)$ at any instant of time $t$ are found to be
\begin{equation}
C_x\left( t\right)  =\left( \frac{f_s^{2}+2\gamma k_{B}T}{2k\beta}\right) e^{ -\frac{\beta t}{2m}} \left( \cos \omega_1 t +\frac{\beta}{2m \omega_{1}} \sin  \omega_1 t\right)
\end{equation}
and
\begin{equation}
C_v\left( t\right)  =\left( \frac{f_s^{2}+2\gamma k_{B}T}{2m\beta}\right) e^{-\frac{\beta t}{2m}} \left( \cos  \omega_1 t -\frac{\beta}{2m \omega_{1}} \sin  \omega_1 t\right).
\end{equation}
Here, $\omega_{1}=\sqrt{\frac{k}{m}-\frac{\beta^{2}}{4m^{2}}}$.

Finally, we have considered the case when the particle is confined in a harmonic well and driven by a sinusoidal driving force $A\sin(\omega t)$, where $A$ and $\omega$ as the amplitude and frequency of the drive, respectively. The dynamics can be described as
\begin{equation}
m\frac{dv}{dt}=-\gamma v-kx+ A\sin \left( \omega t\right)+f_s\left( 1-\frac{v}{v_0}\right)+ \xi\left( t\right).
\end{equation}
By solving the dynamics and following the same procedure as in the above two cases, the power spectral densities are obtained to be
\setlength\belowdisplayskip{0pt}
\begin{equation}
S_x\left( \omega\right)  =\left(\frac{f_{s}^{2}+2 \gamma k_{B}T+\frac{A^{2}}{16}}{\left( m \omega^{2}-k\right) ^{2}+\left( \omega \beta\right) ^{2}}\right)  
\end{equation}
and
\begin{equation}
S_{v}\left( \omega\right) =\omega^{2}\left( \frac{ f_{s}^{2}+2 \gamma k_{B}T+\frac{A^{2}}{16}}{\left( m \omega^{2}-k\right) ^{2}+\left( \omega \beta\right) ^{2}}\right). 
\end{equation}
Similarly, $C_x(t) $ and $C_v(t)$ are calculated to be
\begin{equation}
C_x\left( t\right) =P e^{  -\frac{\beta t}{2m}} \left( \cos  \omega_1 t + \frac{\beta}{2m \omega_{1}} \sin  \omega_1 t\right) 
\end{equation}
 and
\begin{equation}
C_v\left( t\right)  =Q e^{ -\frac{\beta t}{2m}}\left( \cos  \omega_1 t - \frac{\beta}{2m \omega_{1}} \sin  \omega_1 t\right).
\end{equation}
with $P=\left( \frac{f_s^{2}+2\gamma k_{B}T+\frac{A^{2}}{16}}{2k\beta }\right)$ and $Q=\left( \frac{f_s^{2}+2\gamma k_BT+\frac{A^{2}}{16}}{2m\beta}\right)$

In the case of confined harmonic particle, we have also calculated the correlation $C^{T}_{x}$ corresponding to the shift relative to the average position of the particle, which is in the form
\begin{equation}
C^{T}_{x} =\left(\frac{\gamma k_{B}T}{k\beta}\right) e^{ -\frac{\beta t}{2m}} \left( \cos \omega_1 t +\frac{\beta}{2m \omega_{1}} \sin  \omega_1 t\right).
\end{equation}
This is simply the thermal part of the correlation function $C_{x}(t)$ in both Eq.~[9] and Eq.~[14].

\section{\textbf{Results and Discussion}}
In Fig.~1, we have plotted the velocity auto correlation function as a function of time [Eq.~(5)] for different values of $f_{s}$. It is observed that $C_{v}(t)$ decays exponentially with time reflecting that the correlation in velocity decreases with time. In the longer time regime for $f_{s}=0$, $C_{v}(t)$ approaches zero value due to decorrelation. When the particle is subjected to a linear velocity dependent force, $C_{v}(t)$ also decay exponentially with time and gets saturated to a finite value in the time asymptotic regime. This clearly indicates that the correlation between the velocities at two different times still exists even after the system has reached the steady state. The critical time beyond which $C_{v}(t)$ gets saturated, decreases with increase in the $f_s$ value. In the absence of force ($f_{s}=0$), the particle moves freely. When the force is applied, the motion of the particle is controlled and driven in the direction of the force. As a result, the free movement of the particle is restricted. As the value of $f_{s}$ increases, the particle attains the steady state faster leading to a reduction in the critical time. On the other hand, for a fixed $f_s$ value, $C_{v}(t)$ decays faster with $\gamma$.

The nonzero value of the velocity autocorrelation also reflect the equilibrium solution and energetic stability of the particle in the time asymptotic limit. The equilibrium value of course depend on the nature of the velocity dependent force. Similar conclusion has been drawn for a particular case of the model studied in Ref.~[\onlinecite{karmeshu1976motion}]. In their model, the viscous force consists of a deterministic part which represents the intrinsic damping of the medium and a stochastic part which is basically a random velocity dependent force. Further, they have considered two different cases i.e. in the presence and in the absence of a random driving force. In the former case, when the coefficient of the random velocity dependent force is a constant, our model can be treated as a special case of the model used in Ref.~[\onlinecite{karmeshu1976motion}].

\begin{figure}[hbtp]
\centering
\includegraphics[scale=0.6]{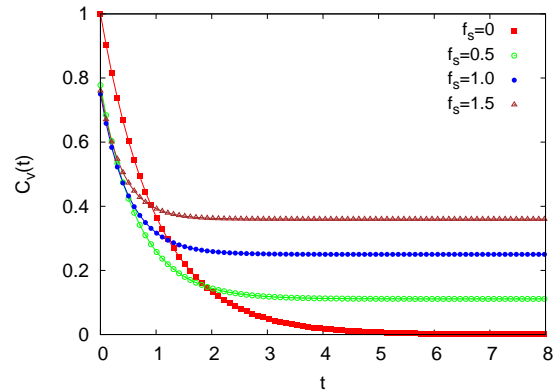}
\caption{$C_v(t)$ [Eq.~(5)] as a function of $t$ for different values of $f_s$. The other parameters ($\gamma$, $v_0$, and $m$) are fixed to $1$}.
\end{figure}

The power spectral density for the position of the particle [$S_x(\omega)$] and the corresponding autocorrelation function [$C_{x}(t)$] for various parameter regimes of the model when the particle is confined in a harmonic well, are presented in Fig.~2. Figures~2(a) and 2(b) show $S_x$ and $C_x$ as a function of $\omega$ and $t$, respectively for various $f_s$ values. The $S_x(\omega)$ spectra for all values of $f_{s}$, exhibit symmetric behaviour with exponential tails on either sides, as in case of a driven harmonic system\cite{SAHOO20086284}. For $f_{s}=0$, the spectrum has a double peaked structure. The optimum values of $\omega$ at which the spectrum $S_{x}(\omega)$ shows peaks are obtained by equating $\frac{dS_{x}(\omega)}{d\omega}$ to zero. For any value of $f_{s}$, the peaks are calculated to be either at $\omega=0$ or at $\omega=\pm \sqrt{\frac{k}{m}-\frac{\beta^{2}}{2m^{2}}}$. For $f_{s}=0$, the peaks are found to be at $\omega=\pm \sqrt{\frac{k}{m}-\frac{\gamma^{2}}{2m^{2}}}$. With increase in $f_{s}$ value, the double peaking behaviour of the spectrum is transformed into a single peak, centered at zero. However, the intensity of the spectrum is found to be reduced initially and then increases for larger $f_{s}$ values as one can notice from Eq. 7. The position autocorrelation function, $C_x(t)$ [Fig.~2(b)] for $f_{s}=0$ shows a non monotonic variation with time. In the lower $t$ range, it decays exponentially, attains a minimum with negative value in the intermediate time range, and then attains the zero value in the long time limit. This showcases the conventional behaviour of the position autocorrelation function\cite{wright2003forward}. The approach of $C_{x}(t)$ to zero reflects that the position at different times get decorrelated in the time asymptotic limit ($t \rightarrow \infty$). As $f_{s}$ increases, the minimum gradually disappear and the variation of $C_{x}$ becomes more and more monotonic with $t$. In the long time limit, $C_{x}(t)$ approaches zero much faster than the case of $f_{s}=0$. This is because, as the value of $f_{s}$ increases the particle mostly follow the direction of force and the probability of changing it's position inside the harmonic well decreases. As a result of this, the non monotonic behaviour of $C_x(t)$ gets suppressed and approaches zero comparatively faster than the small $f_{s}$ values. 

Figures~ 2(c) and 2(d) depict $S_{x}(\omega)$ and $C_{x}(t)$, respectively for different values of $\gamma$. In Fig.~2(c), it is observed that the peak value of the $S_{x}(\omega)$ spectrum increases and the spectrum becomes sharper with increase in $\gamma$ value. The corresponding $C_x(t)$ in Fig.~2(d) decays with time and does not show any systematic behaviour with $\gamma$. The behaviour of the position correlation with $\gamma$ is not well understood because of the complex interplay of the dynamics. Similarly, $S_{x}(\omega)$ and $C_{x}(t)$ are calculated for different values of $k$ and are presented in Fig.~2(e) and 2(f), respectively. The peak value in the $S_{x}(\omega)$ spectrum decreases and the spectrum becomes broader with increase in $k$ [Fig.~2(e)] which is opposite to the behaviour observed for varying $\gamma$ values. The corresponding $C_x(t)$ in Fig.~2(f) shows an exponentially decaying behaviour with time and becomes zero in the long time limit. With increase in the value of $k$, the variation of $C_{x}(t)$ with $t$ becomes more and more non-monotonic and approaches a conventional damped oscillatory behaviour in the larger $k$ limit. As the value of $k$ increases, the particle becomes more and more confined in the well and the random fluctuations of the particle across the mean position of the harmonic well increases. As a result, the fluctuations in the spectrum increases with increase in $k$, thereby, suppressing the $S_{x}(\omega)$ spectral intensity and producing a damped oscillatory behaviour of $C_{x}(t)$. 
 \begin{figure}[htbp]
\includegraphics[scale =0.7]{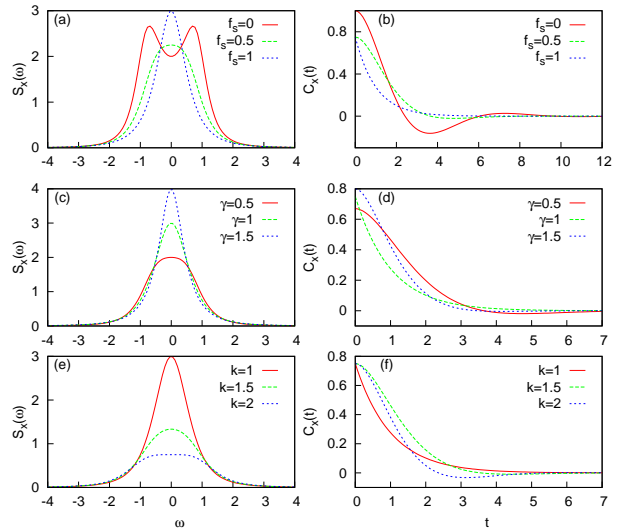}
\caption{For different values of $f_{s}$, (a) $S_{x}$ vs $\omega$ [Eq.~(7)] and (b) $C_{x}$ vs $t$ [Eq.~(9)], fixing $\gamma=k=1$. For different values of $\gamma$, (c) $S_{x}$ vs $\omega$ [Eq.~(7)] and (d) $C_{x}$ vs $t$ [Eq.~(9)], fixing $f_{s}=k=1$. For different values of $k$, (e) $S_{x}$ vs $\omega$ [Eq.~(7)]and (f) $C_{x}$ vs $t$ [Eq.~(9)], fixing $f_{s}=\gamma=1$. The other common fixed parameters are $v_0=m=1$}
\end{figure} 

The spectral density for the velocity of the particle [$S_{v}(\omega)$] and the corresponding $C_{v}(t)$ in various parameter regimes of the model when the particle is confined in a harmonic well, are plotted in Fig.~3. In Figs.~3(a) and 3(b), $S_{v}(\omega)$ and the corresponding $C_v(t)$ are shown for different values of $f_{s}$. The symmetric nature of $S_{v}(\omega)$ resembles the classical behaviour as expected \cite{wright2003forward}. By equating $\frac{dS_{v}(\omega)}{d\omega}$ to zero, the peak positions in the $S_{v}(\omega)$ spectrum are obtained to be at $\omega=\pm \sqrt{\frac{k}{m}}$ and hence are independent of $f_{s}$ values. For this reason, the peaks in the spectrum appear at the same $\omega$ irrespective of the change in $f_{s}$ values. However, the peak heights in the $S_{v}(\omega)$ spectrum get suppressed with increase in $f_{s}$ and the corresponding $C_v(t)$ decays exponentially with $t$. For $f_{s}=0$, the nature of the curve follows the damped oscillating behaviour as expected classically\cite{wright2003forward}. In the absence of the linear velocity dependent force, the particle is confined in a harmonic trap and oscillates randomly across the mean position of the well, making both forward and backward movements frequently. Therefore, $C_{v}(t)$ oscillates between the positive and negative values and finally becomes zero in the $t\rightarrow\infty$ limit. With increase in $f_{s}$, the particle is dragged and follows the direction of the linear velocity dependent force. As a result, the harmonic effect is dominated and the oscillatory behaviour of the particle gets suppressed for larger $f_{s}$ values. This leads to faster decay of $C_{v}(t)$.

In Figs.~3(c) and 3(d), we have plotted $S_v(\omega)$ and the corresponding $C_v(t)$ for different values of $\gamma$. We observe that with increase in $\gamma$, the peak in the $S_v(\omega)$ spectrum is suppressed and the spectrum becomes broader. However, the peak positions remain unchanged with varying $\gamma$. The corresponding $C_{v}(t)$ decays with $t$ and approaches zero in the time asymptotic limit. A weak damped oscillatory behavior is also observed for $C_{v}(t)$ but the variation of $C_v(t)$ with $\gamma$ is found to be not systematic. It is expected that with increase in $\gamma$, the motion of the particle should be obstructed by the presence of viscous drag which may reduce the harmonic oscillatory behaviour. However, in the present case, the non-systematic variation of $C_{v}(t)$ with $\gamma$ implies a complex interplay of the dynamics and reflects it's non-equilibrium nature.

In Figs.~3(e) and 3(f), we have plotted $S_v(\omega)$ and the corresponding $C_v(t)$ for different values of $k$. It is observed that the peaks in the $S_{v}(\omega)$ spectrum shift towards the +ve and -ve direction of $\omega$ values, respectively with increase in $k$, as expected. The corresponding $C_v(t)$ shows a prominent damped oscillatory behaviour with increase in $k$. This is due to the fact that the harmonic confinement becomes stronger with increase in $k$. The probability of oscillatory behaviour of motion of the particle across the mean position of the well increases, reversing it's direction of motion frequently \cite{wright2003forward}.
\begin{figure}[htbp]
 \centering
\includegraphics[scale =0.75]{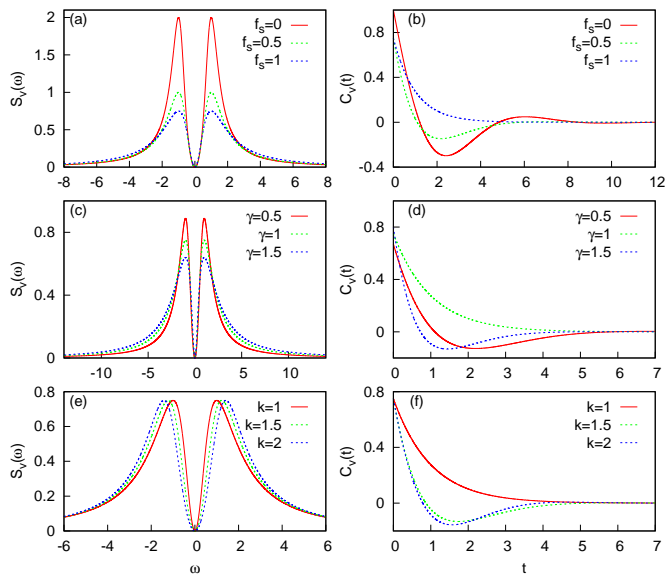}
\caption{For different values of $f_{s}$, (a) $S_{v}$ vs $\omega$ [Eq.~(8)] and (b) $C_{v}$ vs $t$ [Eq.~(10)], fixing $\gamma=k=1$. For different values of $\gamma$, (c) $S_{v}$ vs $\omega$ [Eq.~(8)] and (d) $C_{v}$ vs $t$ [Eq.~(10)], fixing $f_{s}=k=1$. For different values of $k$, (e) $S_{v}$ vs $\omega$ [Eq.~(8)] and (f) $C_{v}$ vs $t$ [Eq.~(10)], fixing $f_{s}=\gamma=1$. The other common fixed parameters are $v_0=m=1$.}
\end{figure}

\begin{figure}[htbp]
 \centering
\includegraphics[scale =0.69]{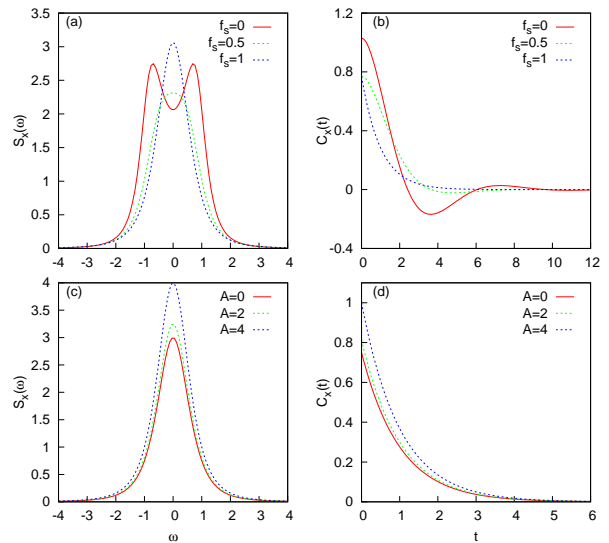}
\caption{For different values of $f_{s}$, (a) $S_{x}$ vs $\omega$ [Eq.~(12)] and (b) $C_{x}$ vs $t$ [Eq.~(14)], fixing $\gamma=k=A=1$. For different values of $A$, (c) $S_{v}$ vs $\omega$ [Eq.~(12)] and (d) $C_{v}$ vs $t$ [Eq.~(14)], fixing $k=\gamma=f_{s}=1$. The other common fixed parameters are $v_0=m=\gamma=k=1$.}
\end{figure}
Figure~4 presents $S_{x}$ vs $\omega$ and the corresponding $C_{x}$ vs $t$ for different values of $f_{s}$ and in different parameter regimes of the model, when the particle is confined in a harmonic well and is subjected to an external driving force, $A\sin(\omega t)$. The behaviour of $S_{x}(\omega)$  [Fig.~4(a)] and $C_{x}(t)$ [Fig.~4(b)] are similar to the behaviour observed in Figs.~2(a) and 2(b) in the absence of the sinusoidal driving force. The intensity of the $S_x(\omega)$ spectrum [Fig.~4(c)] is found to increase with increase in amplitude of drive ($A$). 
The corresponding $C_{x}(t)$ [Fig.~4(d)] decays exponentially with time and in the transient regime, the magnitude of $C_{x}(t)$ is found to increase with increase in $A$. We have also studied $S_{x}(\omega)$ and $C_{x}(t)$ for different values of $\gamma$ and $k$, fixing other parameters to unity, but no significant change in behaviour is observed. 

Finally for a confined harmonic particle, we have exactly calculated the correlation function corresponding to the shift relative to the average position of the particle and is given by the expression in Eq.~16. From this expression, it is confirmed that this term is due to the thermal contribution of the position correlation function and is also $f_{s}$ dependent. This term survives and contributes to the the whole position correlation function $C_{x}(t)$ even in the absence of linear velocity dependent force, i.e., in $f_{s} \rightarrow 0$ limit. However, for the athermal regime, i.e., in $T \rightarrow 0$ limit, the  correlation function is entirely due to the presence of linear velocity dependent force, $f_{s}$.  
\begin{figure}
 \centering
\includegraphics[scale =0.7]{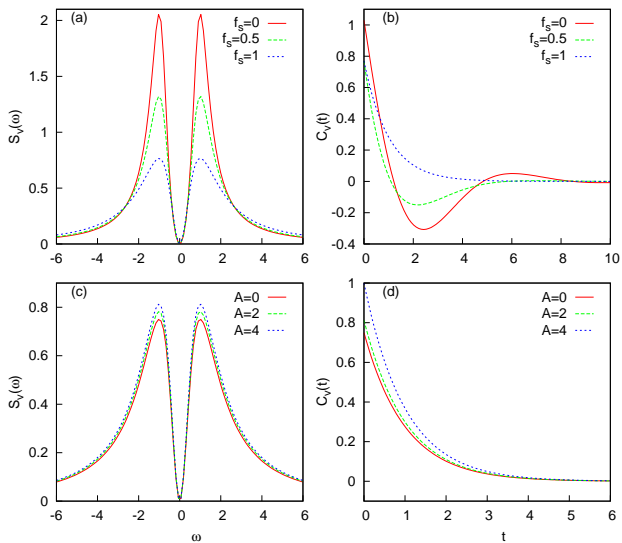}
\caption{For different values of $f_{s}$, (a) $S_{v}$ vs $\omega$ [Eq.~(13)]and (b) $C_{v}$ vs $t$ [Eq.~(15)], fixing $k=\gamma=A=1$. For different values of $A$, (c) $S_{v}$ vs $\omega$ [Eq.~(13)] and (d) $C_{v}$ vs $t$ [Eq.~(15)], fixing $k=\gamma=f_{s}=1$. The other common fixed parameters are $v_0=m=\gamma=k=1$.}
\end{figure}

In Fig.~5, we have plotted $S_v(\omega)$ and the corresponding $C_v(t)$ for different parameter regimes of the model. The overall behaviour of $S_{v}(\omega)$ [Fig.~5(a)] and $C_{v}(t)$ [Fig.~5(b)] are found to be similar to the behaviour observed in Figs.~3(a) and 3(b), respectively for varying $f_{s}$ and in the absence of sinusoidal driving force. The peak in the $S_v(\omega)$ spectrum increases with increase in $A$. The corresponding $C_v(t)$ decays exponentially with time and approaches zero in the time asymptotic limit. At a particular instant of time, the magnitude of $C_v(t)$ increases with increase in $A$. This implies that the correlation between velocities at different times increases with increase in the amplitude of the drive. This is because, the particle while randomly fluctuating across the mean position of the well, takes the advantage of the sinusoidal drive and follows the nature of the driving force, giving a positive contribution to $C_{v}(t)$. The behaviour of $S_{v}(\omega)$
and $C_{v}(t)$ are also studied for varying $k$ and $\gamma$ but no visible changes are observed.

\section{Conclusion}
To understand the dynamical properties of a Brownian particle confined in a harmonic well and in the presence of a linear velocity dependent force, we have exactly calculated and analyzed the position and velocity autocorrelation functions. It is observed that in the presence of a linear velocity dependent force and when the particle is not confined, the velocity autocorrelation function decays exponentially with time and saturates in the time asymptotic limit. The nonzero and constant value in the long time limit clearly indicate that a finite correlation still exists in the time asymptotic limit. This also confirms that the mean energy approaches equilibrium value and provides a stable solution of the particle at the time asymptotic limit, depending on the nature of the velocity dependent force. When the particle is confined in a harmonic well, the velocity autocorrelation function shows a damped oscillatory behaviour before approaching to zero in the long time limit. A non-systematic variation of the correlation is observed with viscous coefficient of the medium. When the particle is further subjected to a sinusoidal driving force, the damped oscillatory behaviour is suppressed, the velocity autocorrelation function decays exponentially with time, and the steady state is approached faster. It is worth mentioning that our results can be applicable to the simple models of molecular motor obeying the linear force velocity relation. This is possible if we assume that the motors are self propelled particles in a Langevin heat bath and the self propulsion mechanism is due to the autonomous linear velocity dependent force, ignoring all other specific details of self propulsion mechanism. 

\section{Acknowledgments}
We would like to thank Arnab Saha for his valuable suggestions. We also thank Debasish Chaudhury for useful discussions during 7th Indian Statistical Physics Community Meeting (Code:ICTS/ispsm2020/02) held at ICTS. MS acknowledges the INSPIRE Faculty award (IFA 13 PH-66) by the Department of Science and Technology and the UGC Faculty recharge program (FRP-56055) for the financial support.\\

\textbf{Contribution of Authors:} MS designed the research problem, developed the model and supervised the complete work. Some initial calculations are done by SS and AN completed all the calculations. AN and MS analyzed the data and MS wrote the final manuscript.
\bibliographystyle{unsrtnat}


\begin{thebibliography}{41}
\providecommand{\natexlab}[1]{#1}
\providecommand{\url}[1]{\texttt{#1}}
\expandafter\ifx\csname urlstyle\endcsname\relax
  \providecommand{\doi}[1]{doi: #1}\else
  \providecommand{\doi}{doi: \begingroup \urlstyle{rm}\Url}\fi

\bibitem[Strogatz()]{strogatz1994nonlinear}
S~.H. Strogatz.
\newblock \emph{Nonlinear Dynamics and Chaos}.
\newblock (Perseus Books, Reading, MA, 1994).

\bibitem[Einstein((2005))]{einstein2005theorie}
A.~Einstein.
\newblock Zur theorie der brownschen bewegung [adp 19, 371 (1906)].
\newblock \emph{Annalen der Physik}, \textbf{14}:\penalty0 248, (2005).

\bibitem[Nelson()]{nelson1967dynamical}
E.~Nelson.
\newblock \emph{Dynamical theories of Brownian motion}.
\newblock (Princeton university press, 1967).

\bibitem[Mazo(2002)]{mazo2002brownian}
R.~M. Mazo.
\newblock \emph{Brownian motion: fluctuations, dynamics, and applications}.
\newblock (Oxford University Press), 2002.

\bibitem[Bechinger((2005))]{babivc2005colloids}
D.~Babi{\v{c}}{,} C. Schmitt{,}~C. Bechinger.
\newblock Colloids as model systems for problems in statistical physics.
\newblock \emph{Chaos}, \textbf{15}:\penalty0 026114, (2005).

\bibitem[Uhlenbeck and Ornstein((1930))]{uhlenbeck1930theory}
G.~E. Uhlenbeck and L.~S. Ornstein.
\newblock On the theory of the brownian motion.
\newblock \emph{Phys. Rev.}, \textbf{36}:\penalty0 823, (1930).

\bibitem[Chow and Hermans((1973))]{CHOW1973156}
T.S. Chow and J.J. Hermans.
\newblock Brownian motion of a spherical particle in a compressible fluid.
\newblock \emph{Physica}, {\textbf{65}}:\penalty0 156, (1973).

\bibitem[Chandrasekhar((1949))]{chandrasekhar1949brownian}
S.~Chandrasekhar.
\newblock Brownian motion, dynamical friction and stellar dynamics.
\newblock \emph{Rev. Mod. Phys.}, \textbf{21}:\penalty0 383, (1949).

\bibitem[Parisi((2005))]{parisi2005brownian}
G.~Parisi.
\newblock Brownian motion.
\newblock \emph{Nature}, \textbf{433}:\penalty0 221, (2005).

\bibitem[Kroy((2005))]{frey2005brownian}
E.~Frey{,}~K. Kroy.
\newblock Brownian motion:a paradigm of soft matter and biological physics.
\newblock \emph{Ann. Phys.}, \textbf{14}:\penalty0 20, (2005).

\bibitem[Astumian((1997))]{Astumian917}
R.~Dean Astumian.
\newblock Thermodynamics and kinetics of a brownian motor.
\newblock \emph{Science}, \textbf{276}:\penalty0 917, (1997).

\bibitem[Mason((2010))]{squires2010fluid}
T.~M. Squires {,} T.~G. Mason.
\newblock Fluid mechanics of microrheology.
\newblock \emph{Annu. Rev. Fluid Mech.}, \textbf{42}:\penalty0 413, (2010).

\bibitem[Wirtz((2009))]{wirtz2009particle}
D.~Wirtz.
\newblock Particle-tracking microrheology of living cells: principles and
  applications.
\newblock \emph{Annu. Rev. Biophys}, \textbf{38}:\penalty0 301, (2009).

\bibitem[Marchesoni((2009))]{hanggi2009artificial}
P.~H{\"a}nggi{,}~F. Marchesoni.
\newblock Artificial brownian motors:controlling transport on the nano scale.
\newblock \emph{Rev. Mod. Phys.}, \textbf{81}:\penalty0 387, (2009).

\bibitem[Langevin(1908)]{langevincomptes}
P~Langevin.
\newblock Comptes rendues \textbf{146},.
\newblock \emph{Google Scholar CAS}, page 530, 1908.

\bibitem[Lis{\`y}((2011))]{tothova2011langevin}
J.~T{\'o}thov{\'a}{,} G. Vasziov{\'a}{,} L. Glod{,}~V. Lis{\`y}.
\newblock A note on 'langevin theory anomalous brownian motion made simple.
\newblock \emph{Eur .J. Phys.}, \textbf{32}:\penalty0 645, (2011).

\bibitem[Natsuki()]{kubo2012statistical}
R.~Kubo{,} M. Toda{,}~H. Natsuki.
\newblock \emph{Nonequilibrium Statistical Mechanics}.
\newblock (Springer,2012).

\bibitem[Golestanian((2007))]{howse2007self}
J.~R. Howse{,} R. A. L. Jones{,} A. J. Ryan{,} T. Gough{,} R. Vafabakhsh{,}~R.
  Golestanian.
\newblock Self-motile colloidal particles:from directed propulsion to random
  walk.
\newblock \emph{Phys .Rev. Lett.}, \textbf{99}:\penalty0 048102, (2007).

\bibitem[Reimann((2002))]{reimann2002brownian}
P.~Reimann.
\newblock Brownian motors: noisy transport far from equilibrium.
\newblock \emph{Phys. Rep.}, \textbf{361}:\penalty0 57, (2002).

\bibitem[Crespi((2004))]{paxton2004catalytic}
W.~F. Paxton{,} K.C. Kistler{,} C. C. Olmeda{,} A. Sen{,} S. K. St. Angelo{,}
  Y. Cao{,} T. E. Mallouk{,} P. E. Lammert{,} V.~H. Crespi.
\newblock Catalytic nanomotors: autonomous movement of striped nanorods.
\newblock \emph{J. Am. Chem. Soc.}, \textbf{126}:\penalty0 13424, (2004).

\bibitem[van Teeffelen {,}~Lowen((2011))]{hagen2011brownian}
B.~Hagen {,}~S. van Teeffelen {,}~Lowen.
\newblock Brownian motion of a self-propelled particle.
\newblock \emph{J. Phys.: Condens. Matter}, \textbf{23}:\penalty0 194119,
  (2011).

\bibitem[Schweitzer()]{schweitzer2003brownian}
F.~Schweitzer.
\newblock \emph{Brownian agents and active particles: collective dynamics in
  the natural and social sciences}.
\newblock (Springer Science \& Business Media 2003).

\bibitem[Tilch((1998))]{schweitzer1998complex}
F.~Schweitzer{,} W. Ebeling{,}~B. Tilch.
\newblock Complex motion of brownian particles with energy depots.
\newblock \emph{Phys. Rev. Lett.}, \textbf{80}:\penalty0 5044, (1998).

\bibitem[Le((2008))]{zhang2008active}
Y.~Zhang{,} C. K. Kim{,} K. J.~B. Le.
\newblock Active motions of brownian particles in a generalized energy-depot
  model.
\newblock \emph{New J. Phys.}, \textbf{10}:\penalty0 103018, (2008).

\bibitem[Howard()]{howard2001mechanics}
J.~Howard.
\newblock \emph{Mechanics of Motor protiens and the Cytoskeleton}.
\newblock (Sinauer Assosiates,MA,2001).

\bibitem[Block((1994))]{savoda1994cell}
K.~Svoboda{,} S.~M. Block.
\newblock \emph{Cell}, \textbf{77}:\penalty0 773, (1994).

\bibitem[Chaudhiri((2013))]{ganguly2013}
C.~Ganguly{,}~D. Chaudhiri.
\newblock Stochastic thermodynamics of active brownian particle.
\newblock \emph{Phys. Rev. E.}, \textbf{88}:\penalty0 032102, (2013).

\bibitem[Volpe((2013))]{volpe2013simulation}
G.~Volpe{,}~G. Volpe.
\newblock Simulation of a brownian particle in an optical trap.
\newblock \emph{Am. J. Phys}, \textbf{81}:\penalty0 224, (2013).

\bibitem[Bellissima et~al.(2015)Bellissima, Neumann, Guarini, Bafile, and
  Barocchi]{Bellissima}
S.~Bellissima, M.~Neumann, E.~Guarini, U.~Bafile, and F.~Barocchi.
\newblock Time dependence of the velocity autocorrelation function of a fluid:
  An eigenmode analysis of dynamical processes.
\newblock \emph{Phys. Rev. E}, \textbf{92}:\penalty0 042166, 2015.

\bibitem[Chakraborty((2011))]{chakraborty2011velocity}
D.~Chakraborty.
\newblock Velocity autocorrelation function of a brownian particle.
\newblock \emph{Eur. Phys. J. B.}, \textbf{83}, (2011).

\bibitem[Williams et~al.(2006)Williams, Bryant, Snook, and van Megen]{Williams}
Stephen~R. Williams, G.~Bryant, I.~K. Snook, and W.~van Megen.
\newblock Velocity autocorrelation functions of hard-sphere fluids: Long-time
  tails upon undercooling.
\newblock \emph{Phys. Rev. Lett.}, \textbf{96}:\penalty0 087801, 2006.

\bibitem[Kim((2012))]{do2012force}
Su~Do Yi{,} B.~J. Kim.
\newblock \emph{Computs}, \textbf{183}:\penalty0 1574, (2012).

\bibitem[Makri((2003))]{wright2003forward}
N.~J. Wright{,}~N. Makri.
\newblock Forward-backward semiclassical dynamics for condensed phase time
  correlation functions.
\newblock \emph{J. Chem. Phys.}, \textbf{119}:\penalty0 1634, (2003).

\bibitem[Cichocki and Felderhof(1995)]{cichocki}
B.~Cichocki and U.~Felderhof.
\newblock Velocity autocorrelation function of interacting brownian particles.
\newblock \emph{Physical review. E, Statistical physics, plasmas, fluids, and
  related interdisciplinary topics}, \textbf{51}:\penalty0 5549, 1995.

\bibitem[Atzberger(2006)]{ATZBERGER2006225}
Paul~J. Atzberger.
\newblock Velocity correlations of a thermally fluctuating brownian particle: A
  novel model of the hydrodynamic coupling.
\newblock \emph{Physics Letters A}, \textbf{351}:\penalty0 225, 2006.

\bibitem[Klein((1981))]{hanna1981velocity}
S.~Hanna{,} W. Hess{,}~R. Klein.
\newblock \emph{J. Phys.}, \textbf{14}:\penalty0 L493, (1981).

\bibitem[Chandrasekhar((1943))]{chandrasekhar1943stochastic}
S.~Chandrasekhar.
\newblock Stochastic problems in physics and astronomy.
\newblock \emph{Rev. Mod. Phys.}, \textbf{15}:\penalty0 1, (1943).

\bibitem[Sahoo.((2007))]{jayannavar2007charged}
A.~M. Jayannavar{,}~M. Sahoo.
\newblock A charged particle in a magnetic field:jarzynski equality.
\newblock \emph{Phys. Rev. E.}, \textbf{75}:\penalty0 032102, (2007).

\bibitem[Sahoo((2019))]{sahoo2019transport}
M~Sahoo.
\newblock Transport coherence and diffussion in a temporal asymmetric rocked
  ratchet model.
\newblock \emph{Int. J. Mod. Phys. B.}, \textbf{33}:\penalty0 1950096, (2019).

\bibitem[Karmeshu((1976))]{karmeshu1976motion}
Karmeshu.
\newblock \emph{J. Appl. Probab.}, \textbf{13}:\penalty0 684, (1976).

\bibitem[Sahoo et~al.((2008))Sahoo, Saikia, Mahato, and
  Jayannavar]{SAHOO20086284}
M.~Sahoo, S.~Saikia, M~C. Mahato, and A.~M. Jayannavar.
\newblock Stochastic resonance and heat fluctuations in a driven double-well
  system.
\newblock \emph{Physica A: Statistical Mechanics and its Applications},
  \textbf387:\penalty0 6284, (2008).
\end{thebibliography}
\end{document}